\documentclass[10pt]{iopart}
\usepackage{iopams}
\usepackage{epsfig}
\usepackage{amssymb}
\usepackage{mathrsfs}

\newcommand{\calA}{{\mathcal A}}
\newcommand{\calC}{{\mathcal C}}
\newcommand{\calP}{{\mathcal P}}
\newcommand\Laplace{\nabla^2_\perp}

\newcommand{\rrr}{\vec{r}}
\newcommand{\vvv}{\vec{v}}
\newcommand{\EEE}{\vec{E}}

\newcommand{\rrrp}{\vec{r}_\perp}
\newcommand{\nablabf}{\vec{\nabla}}

\newcommand{\ex}{\vec{e}_x}
\newcommand{\ey}{\vec{e}_y}
\newcommand{\ez}{\vec{e}_z}

\begin{document}

\title[Mortensen, Olesen \& Bruus:
Transport coefficients for electrolytes in ...]{Transport
coefficients for electrolytes in arbitrarily shaped nano and
micro-fluidic channels}

\author{N.~A. Mortensen\footnote[3]{Corresponding author: nam@mic.dtu.dk},
L.~H. Olesen, and H. Bruus}

\address{MIC -- Department of Micro and Nanotechnology, NanoDTU,
Technical University of Denmark, Bld. 345 east, DK-2800 Kongens Lyngby, Denmark}

\begin{abstract}
We consider laminar flow of incompressible electrolytes in long,
straight channels driven by pressure and electro-osmosis. We use a
Hilbert space eigenfunction expansion to address the general
problem of an arbitrary cross section and obtain general results
in linear-response theory for the hydraulic and electrical
transport coefficients which satisfy Onsager relations. In the
limit of non-overlapping Debye layers the transport coefficients
are simply expressed in terms of parameters of the electrolyte as
well as the geometrical correction factor for the
Hagen--Poiseuille part of the problem. In particular, we consider
the limits of thin non-overlapping as well as strongly overlapping
Debye layers, respectively, and calculate the corrections to the
hydraulic resistance due to electro-hydrodynamic interactions.
\end{abstract}


\submitto{\NJP (accepted)}

 \maketitle

\section{Introduction}

Laminar Hagen--Poiseuille and electro-osmotic flow is important to
a variety of lab-on-a-chip applications and
microfluidics~\cite{Laser:04,Stone:04a,Squires:05a} and the rapid
development of micro and nano fabrication techniques during the
past decade has put even more emphasis on flow in channels with a
variety of shapes depending on the fabrication technique in use.
The list of examples includes rectangular channels obtained by hot
embossing in polymer wafers, semi-circular channels in
isotropically etched surfaces, triangular channels in KOH-etched
silicon crystals, Gaussian-shaped channels in laser-ablated
polymer films, and elliptic channels in stretched PDMS
devices~\cite{Geschke:04a}. While general results for the
shape-dependence of the hydraulic resistance in the case of a
non-conducting fluid were reported recently~\cite{Mortensen:05b}
there has, according to our knowledge, been no analogous detailed
study of the shape-dependence of flow of electrolytes in the
presence of a zeta potential which is a scenario of key importance
to lab-on-a-chip applications involving biological liquids/samples
in both
microfluidic~\cite{Schasfoort:1999,Takamura:03,Reichmuth:03} and
nanofluidic
channels~\cite{Daiguji:2004,Stein:2004,Vanderheyden:2005,Brask:05a,Yao:03a,Yao:03b}.

\begin{figure}[b]
\begin{center}
\epsfig{file=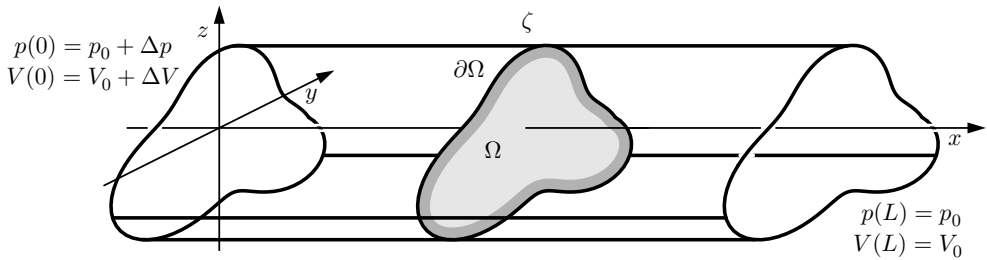, width=\columnwidth, clip}
\end{center}
\caption{A translation invariant channel of arbitrary cross
section $\Omega$ of area $\calA$ containing an electrolyte
driven by a pressure
gradient $-\Delta p/L$ and by electro-osmosis through the
potential gradient $-\Delta V/L$. The channel wall $\partial
\Omega$ has the electrical potential $\zeta$, which induces a
thin, charged Debye layer (dark gray) that surrounds the charge
neutral bulk (light gray). \label{fig1} }
\end{figure}

In this work we address the general steady-state flow problem in
Fig.~\ref{fig1} where pressure gradients and electro-osmosis (EO)
are playing in concert~\cite{ajdari:04a}. We consider a long,
straight channel of length $L$ having a constant cross section
$\Omega$ of area $\calA$ and boundary $\partial\Omega$ of length
$\calP$. The channel contains an incompressible electrolyte, which
we for simplicity assume to be binary and symmetric, i.e.,
containing ions of charge $+Ze$ and $-Ze$ and equal diffusivities $D$.
The electrolyte has the
Debye screening length $\lambda_D$, bulk conductivity $\sigma_\mathrm{\!o}$,
viscosity $\eta$, permittivity $\epsilon$ and at the boundary
$\partial\Omega$ it has a zeta potential $\zeta$. The laminar,
steady-state flow is driven by a linear pressure drop $\Delta p$
and a linear voltage drop $\Delta V$. With these definitions flow
will be in the positive $x$ direction. In the linear-response
regime the corresponding volume flow rate $Q$ and charge current
$I$ are related to the driving fields by
 \begin{equation}\label{eq:G}
 \left(\begin{array}{cc} Q \\ I
 \end{array}\right) =G\left(\begin{array}{cc}
 \Delta p \\ \Delta V
 \end{array}\right),\quad
 G=\left(\begin{array}{cc}G_{11}&G_{12}\\G_{21}&G_{22} \end{array}\right),
 \end{equation}
where, according to Onsager relations~\cite{Brunet:2004}, $G$ is a
symmetric, $G_{12}=G_{21}$, two-by-two conductance matrix. The
upper diagonal element is the hydraulic conductance or inverse
hydraulic resistance given by
 \begin{equation}\label{eq:G11_alpha}
 G_{11} = \frac{\calA}{\alpha\eta}\:\frac{\calA}{L},
 \end{equation}
where $\alpha$ is the dimensionless geometrical correction factor,
shown in Ref.~\cite{Mortensen:05b} to be a linear function of the
dimensionless compactness parameter $\calC=\calP^2/\calA$. While
there is no intrinsic length scale influencing $G_{11}$, the other
elements of $G$ depend on the Debye screening length $\lambda_D$.
This length can be comparable to and even exceed the transverse
dimensions in
nano-channels~\cite{Daiguji:2004,Stein:2004,Vanderheyden:2005}, in
which case the off-diagonal elements may depend strongly on the
actual cross-sectional geometry. However, for thin Debye layers
with a vanishing overlap all four matrix elements in $G$ are
independent of the details of the geometry. For a free
electro-osmotic flow, a constant velocity field
$v^{{}}_\mathrm{eo} = (\epsilon \zeta/\eta)\Delta V/L$ is
established throughout the channel, except for in the thin Debye
layer of vanishing width. Hence $Q = v^{{}}_\mathrm{eo}\calA$ and
 \begin{equation}
 G_{12}= G_{21} =- \frac{\epsilon\zeta}{\eta}\:\frac{\calA}{L},
 \quad \lambda_D \ll \frac{\calA}{\calP}.
 \label{eq:G12thin}
 \end{equation}
From Ohm's law $I = (\sigma_\mathrm{\!o}\calA/L)\Delta V$ it follows that
 \begin{equation}
 G_{22} = \sigma_\mathrm{\!o}\frac{\calA}{L},
 \quad \lambda_D \ll \frac{\calA}{\calP}.
 \label{eq:G22thin}
 \end{equation}
For strongly overlapping Debye layers we find, see Sections
\ref{sec:DH_strongoverlap} and \ref{sec:nonlin_strongoverlap} that
 \begin{eqnarray}
 &G_{12}= G_{21} = -\frac{\epsilon\zeta}{\eta}
 \frac{\sinh\!\Big(\frac{Ze\zeta}{k_BT}\Big)}{\frac{Ze\zeta}{k_BT}}
 \frac{\calA}{\alpha\lambda_D^2}\:\frac{\calA}{L}, &
 \; \lambda_D\gg \frac{\calA}{\calP},
 \label{eq:G12strong}\\
 &G_{22} = \Bigg[\cosh\!\Big(\frac{Ze\zeta}{k_BT}\Big)
 +\frac{\epsilon\zeta^2}{\eta D}
 \frac{\sinh^2\!\Big(\frac{Ze\zeta}{k_BT}\Big)}{\Big(\frac{Ze\zeta}{k_BT}\Big)^2}
 \frac{\calA}{\alpha\lambda_D^2}\Bigg]\:\sigma_\mathrm{\!o}\frac{\calA}{L}
 ,& \; \lambda_D\gg \frac{\calA}{\calP}.
 \label{eq:G22strong}
 \end{eqnarray}
We emphasize that the above results are generally valid for
symmetric electrolytes, even beyond the Debye--H\"uckel
approximation. In the Debye--H\"uckel limit $Ze\zeta\ll k_BT$ they
also hold for asymmetric electrolytes. We also note that in the
Debye--H\"uckel limit the expressions agree fully with the
corresponding limits for a circular cross section and the infinite
parallel plate system, were explicit solutions exist in the
Debye--H\"uckel limit in terms of Bessel
functions~\cite{Rice:65,Probstein:94a} and cosine hyperbolic
functions~\cite{Probstein:94a}, respectively.
From the corresponding resistance matrix $R=G^{-1}$ we get the
hydraulic resistance
 \begin{equation}
 R_{11}= \frac{\alpha}{1-\beta}\frac{\eta L}{\calA^2},
 \end{equation}
where $\beta\equiv G_{12}G_{21}/(G_{11}G_{22})$ is the Debye-layer
correction factor to the hydraulic resistance. In the two limits
we have
 \begin{equation}
 \beta\simeq\alpha\frac{\epsilon^2\zeta^2}{\eta\sigma_\mathrm{\!o} \calA }\times
 \left\{\begin{array}{ccc}1&,&\displaystyle\lambda_D\ll
 \frac{\calA}{\calP}\\
 \\ \displaystyle\Big(\frac{Ze\zeta}{k_BT}\Big)^{-1}
 \sinh\!\Big(\frac{Ze\zeta}{k_BT}\Big)
 \Big(\frac{\calA}{\alpha\lambda_D^2}\Big)^2 &,&\displaystyle\lambda_D\gg
 \frac{\calA}{\calP}\end{array}\right.
 \end{equation}
For $\zeta$ going to zero $\beta$ vanishes and we recover the
result in Ref.~\cite{Mortensen:05b} for the hydraulic resistance.

\section{Governing equations}

For the system illustrated in Fig.~\ref{fig1}, an external
pressure gradient $\nablabf p = -(\Delta p/L)\ex$ and an external
electrical field $\EEE=E \ex=(\Delta V/L)\ex$ is applied. There is
full translation invariance along the $x$ axis, from which it
follows that the velocity field is of the form
$\vvv(\rrr)=v(\rrrp)\ex$ where $\rrrp = y\ey+z\ez$. For the
equilibrium potential and the corresponding charge density we have
$\phi_\mathrm{eq}(\rrr)=\phi_\mathrm{eq}(\rrrp)$ and
$\rho_\mathrm{eq}^e(\rrr)=\rho_\mathrm{eq}^e(\rrrp)$,
respectively. We follow our related recent
work~\cite{Mortensen:05d} and use the Dirac \emph{bra-ket}
notation~\cite{Dirac:81,Merzbacher:70}, where functions $f(\rrrp)$
in $\Omega$ are written as $\big|f\big>$ with inner products
defined by the cross-section integral
 \begin{equation}\label{eq:inner}
 \big< f \big|g\big>\equiv \int_\Omega d\rrrp\, f(\rrrp)g(\rrrp).
 \end{equation}
From the Navier--Stokes equation it follows that the velocity is
governed by~\cite{Batchelor:67,Landau:87a}
 \begin{equation}\label{eq:NS}
 0=\frac{\Delta p}{L}\big|1\big>+\eta \Laplace
 \big|v\big>+\frac{\Delta V}{L} \big|\rho_\mathrm{eq}^e\big>,
\end{equation}
where $\Laplace = \partial^2_y + \partial^2_z$ is the 2D
Laplacian. The equilibrium potential $\big|\phi_\mathrm{eq}\big>$
and the charge density $\big|\rho_\mathrm{eq}^e\big>$ are related
by the Poisson equation
 \begin{equation}\label{eq:P}
 \Laplace \big|\phi_\mathrm{eq}\big> =-
 \frac{1}{\epsilon}\big|\rho_\mathrm{eq}^e \big>.
 \end{equation}
The velocity $\big|v\big>$ is subject to a no-slip boundary
condition on $\partial\Omega$ while the equilibrium potential
$\big|\phi_\mathrm{eq}\big>$ equals the zeta potential $\zeta$ on
$\partial\Omega$. Obviously, we also need a statistical model for
the electrolyte, and in the subsequent sections we will use the
Boltzmann model where the equilibrium potential
$\big|\phi_\mathrm{eq}\big>$ is governed by the Poisson--Boltzmann
equation. However, before turning to a specific model we will
first derive general results which are independent of the
description of the electrolyte.

We first note that because Eq.~(\ref{eq:NS}) is linear we can
decompose the velocity as $\big|v\big> =
\big|v_p\big>+\big|v_\mathrm{eo}\big>$, where $\big|v_p\big>$ is
the Hagen--Poiseuille pressure driven velocity governed by
 \begin{equation}
 0=\frac{\Delta p}{L}\big|1\big>+\eta \Laplace
 \big|v_p\big>,
 \end{equation}
and $\big|v_\mathrm{eo}\big>$ is the electro-osmotic velocity
given by
 \begin{equation}\label{eq:veo}
 \big|v_\mathrm{eo}\big> = -\frac{\epsilon\Delta V}{\eta L}
 \big(\zeta\big|1\big>-\big|\phi_\mathrm{eq}\big>\big).
 \end{equation}
The latter result is obtained by substituting Eq.~(\ref{eq:P}) for
$\big|\rho_\mathrm{eq}^e\big>$ in Eq.~(\ref{eq:NS}). The upper
diagonal element in $G$ is given by $G_{11}=\big<1\big|v_p\big>/\Delta p$
which may be parameterized according to Eq.~(\ref{eq:G11_alpha}).
The upper off-diagonal element is given by
$G_{12}=\big<1\big|v_\mathrm{eo}\big>/\Delta V$ and combined with
the Onsager relation we get
\begin{equation}\label{eq:G12_general}
G_{12}=G_{21}=-\frac{1}{L}\frac{\epsilon}{\eta}\big<1\big|\zeta-\phi_\mathrm{eq}\big>=-\frac{\calA}{L}\frac{\epsilon}{\eta}\big(\zeta-\bar\phi_\mathrm{eq}\big),
\end{equation}
where we have used that $\big< 1\big|1\big>=\calA$ and introduced
the average potential $\bar\phi_\mathrm{eq}=\big<
\phi_\mathrm{eq}\big|1\big>/\big<1\big|1\big>$.

There are two contributions to the lower diagonal element
$G_{22}$; one from migration,
$G_{22}^\mathrm{mig}=\big<1\big|\sigma\big>/L$, and one from
electro-osmotic convection of charge, $G_{22}^\mathrm{conv}=\big<
\rho_\mathrm{eq}^e\big|v_{\rm eo}\big>/\Delta V$, so that
\begin{equation}\label{eq:G22_general}
G_{22}=G_{22}^\mathrm{mig}+G_{22}^\mathrm{conv}=
\frac{1}{L}\big<1\big|\sigma\big>-\frac{\epsilon}{\eta L} \big<
\rho_\mathrm{eq}^e\big|\zeta-\phi_\mathrm{eq}\big>,
\end{equation}
where the electrical conductivity $\sigma(\rrrp)$ depends on the
particular model for the electrolyte. For thin non-overlapping
Debye layers we note that $\bar\phi_\mathrm{eq}\simeq 0$ so that
Eq.~(\ref{eq:G12_general}) reduces to Eq.~(\ref{eq:G12thin}) and,
similarly since the induced charge density is low,
Eq.~(\ref{eq:G22_general}) reduces to Eq.~(\ref{eq:G22thin}). For
strongly overlapping Debye layers the weak screening means that
$\phi_\mathrm{eq}$ approaches $\zeta$ so that the off-diagonal
elements $G_{12}=G_{21}$ and the $G_{22}^\mathrm{conv}$ part of
$G_{22}$ vanish entirely. In the following we consider a
particular model for the electrolyte and calculate the asymptotic
suppression as a function of the Debye screening length
$\lambda_D$.

\section{Debye--H\"uckel approximation}

In the Debye--H\"uckel approximation the equilibrium potential
$\big|\phi_\mathrm{eq}\big>$ is governed by the linearized
Poisson--Boltzmann equation~\cite{Squires:05a}
 \begin{equation}\label{eq:PB_DH}
 \Laplace \big|\phi_\mathrm{eq}\big> =
 \frac{1}{\lambda_D^2}\big|\phi_\mathrm{eq}\big>,
 \end{equation}
where $\lambda_D$ is the Debye screening length. The validity of
this model will be discussed in more detail in
Sec.~\ref{sec:BeyondDebyeHuckel}.

\subsection{Hilbert space formulation}
In order to solve Eqs.~(\ref{eq:NS}), (\ref{eq:P}), and
(\ref{eq:PB_DH}) we will take advantage of the Hilbert space
formulation~\cite{Morse:1953}, often employed in quantum
mechanics~\cite{Merzbacher:70}, and recently employed by us on the
problem of an accelerating Poiseuille flow~\cite{Mortensen:05d}.
The Hilbert space of real functions on $\Omega$ is defined by the
inner product in Eq.~(\ref{eq:inner}) and a complete, countable
set $\big\{\big|\psi_n\big>\big\}$ of orthonormal basis functions,
i.e.,
 \begin{equation}
 \big<\psi_m\big|\psi_n\big>=\delta_{nm},
 \end{equation}
where $\delta_{nm}$ is the Kronecker delta. We choose the
eigenfunctions $\big\{\big|\psi_n\big>\big\}$ of the Helmholtz equation
(with a zero Dirichlet boundary condition on $\partial\Omega$) as
our basis functions,
 \begin{equation} \label{eq:EigenvalDef}
 -\Laplace \big|\psi_n\big> = \kappa_n^2 \big|\psi_n\big>,
 \quad n=1,2,3,\ldots.
 \end{equation}
With this complete basis any function in the Hilbert space can be
written as a linear combination of basis functions. In the
following we write the fields as
 \begin{eqnarray}
 \big|v\big>&=\sum_{n=1}^\infty a_n \big|\psi_n\big>,\label{eq:exp_v} \\
 \big|\phi_\mathrm{eq}\big>&=\zeta\big|1\big>-\sum_{n=1}^\infty b_n
 \big|\psi_n\big>,\label{eq:exp_phi}\\
 \big|\rho_\mathrm{eq}^e\big>&= \sum_{n=1}^\infty c_n
 \big|\psi_n\big>.\label{eq:exp_rho}
 \end{eqnarray}
Inserting Eqs.~(\ref{eq:EigenvalDef}) and~(\ref{eq:exp_phi}) into
Eq.~(\ref{eq:PB_DH}), and multiplying by $\big<\psi_m\big|$,
yields
 \begin{equation} \label{eq:bn}
 b_n = \zeta
 \frac{\big<\psi_n\big|1\big>}{1+(\kappa_n\lambda_D)^2},
 \quad n=1,2,3,\ldots
 \end{equation}
Likewise, inserting Eqs.~(\ref{eq:EigenvalDef}),
(\ref{eq:exp_phi}), (\ref{eq:exp_rho}) and~(\ref{eq:bn}) into
Eq.~(\ref{eq:P}), and multiplying by $\big<\psi_m\big|$, yields
 \begin{equation} \label{eq:cn}
 c_n =-\epsilon\zeta\kappa_n^2
 \frac{\big<\psi_n\big|1\big>}{1+(\kappa_n\lambda_D)^2},
 \quad n=1,2,3,\ldots
 \end{equation}
Finally, using Eqs.~(\ref{eq:exp_v}), (\ref{eq:exp_phi}),
(\ref{eq:bn}) and~(\ref{eq:cn}) in Eq.~(\ref{eq:NS}) leads to
 \begin{equation} \label{eq:an}
 a_n =\left(\frac{\Delta p}{\eta L}\frac{1}{\kappa_n^2}
 -\frac{\epsilon\zeta\Delta V}{\eta L}
 \frac{1}{1+(\kappa_n\lambda_D)^2}\right)\big<\psi_n\big|1\big>,
 \quad n=1,2,3,\ldots
 \end{equation}

\subsection{Transport coefficients}

The flow rate and the electrical current are conveniently written
as
 \begin{eqnarray}
 Q &=\big<1\big|v\big>,\label{eq:Q}\\
 I &= \big<\rho_\mathrm{eq}^e\big|v\big>+\sigma_\mathrm{\!o} E
 \big<1\big|1\big>,\label{eq:I}
 \end{eqnarray}
where the second relation is the linearized Nernst--Planck
equation with the first term being the convection/streaming
current while the second is the ohmic current. Substituting
Eqs.~(\ref{eq:exp_v}) and (\ref{eq:exp_rho}) into these
expressions we identify the transport coefficients as
 \begin{eqnarray}\label{eq:Ggeneral}
 G_{11}&= \frac{\calA}{\eta L}\sum_{n=1}^\infty \frac{1}{\kappa_n^2 }
   \frac{\calA_n}{\cal A},\\
 G_{12} &=-\frac{\epsilon\zeta\calA}{\eta L}
   \sum_{n=1}^\infty \frac{1}{1+(\kappa_n\lambda_D)^2}
   \frac{\calA_n}{\cal A},\label{eq:G12general} \\
 G_{21} &=-\frac{\epsilon\zeta\calA}{\eta L}
   \sum_{n=1}^\infty \frac{1}{1+(\kappa_n\lambda_D)^2}
   \frac{\calA_n}{\cal A},\\
 G_{22}&=\frac{\sigma_\mathrm{\!o}\calA}{L}+
   \frac{(\epsilon\zeta)^2}{\eta\lambda_D^2}
   \frac{\calA}{L}\sum_{n=1}^\infty
   \frac{(\kappa_n\lambda_D)^2}{\big[1+(\kappa_n\lambda_D)^2\big]^2}
   \frac{\calA_n}{\cal A}, \label{eq:Ggeneral:4}
 \end{eqnarray}
where
 \begin{equation}
 \calA_n\equiv
 \frac{\big|\big<1\big|\psi_n\big>\big|^2}{\big<\psi_n\big|\psi_n\big>}
 =\big|\big<1\big|\psi_n\big>\big|^2
 \end{equation}
is the effective area of the eigenfunction $\big|\psi_n\big>$. The
fraction $\calA_n/\calA$ is consequently a measure of the relative
area occupied by $\big|\psi_n\big>$ satisfying the sum-rule
$\sum_{n=1}^\infty \calA_n = \calA$~\cite{Mortensen:05d}. We note
that as expected $G$ obeys the Onsager relation $G_{12}=G_{21}$.
Furthermore, using that
 \begin{equation}
 \frac{(\kappa_n\lambda_D)^2}{\big[1+(\kappa_n\lambda_D)^2\big]^2}
 = -\frac{\lambda_D}{2}
 \frac{\partial}{\partial\lambda_D}\frac{1}{1+(\kappa_n\lambda_D)^2},
 \end{equation}
we get the following bound between the off-diagonal elements
$G_{12}=G_{21}$ and the lower diagonal element $G_{22}$,
 \begin{equation}\label{eq:G22_G12}
 G_{22}=\frac{\sigma_\mathrm{\!o}\calA}{L}+ \frac{\epsilon\zeta}{2\lambda_D}
 \frac{\partial G_{12}}{\partial\lambda_D}.
 \end{equation}

In the context of the geometrical correction factor $\alpha$
studied in detail in~\cite{Mortensen:05b} we note that the first
diagonal element may be written as $G_{11} = G_{11}^*/\alpha$
where $G_{11}^*=\frac{\calA^2}{\eta L}$ is a characteristic
hydraulic conductance and the geometrical correction factor
$\alpha$ can be expressed as~\cite{Mortensen:05d}
\begin{equation}\label{eq:alpha}
\alpha \equiv\left(\sum_{n=1}^\infty \frac{1}{\kappa_n^2 \calA}
\frac{\calA_n}{\cal A}\right)^{-1}=\calC \left(\sum_{n=1}^\infty
\frac{1}{k_n^2} \frac{\calA_n}{\cal A}\right)^{-1},
\end{equation}
where $k_n=\kappa_n \calA/\calP$ is a dimensionless eigenvalue. In
passing we furthermore note that this formal result is a
convenient starting point for perturbative analysis of the
correction due to small changes in the boundary
$\partial\Omega$~\cite{Parker:1998}.

\begin{table}[t!]
\begin{center}
\begin{tabular}{lcccccc}
& $k_1^2$ & $\calA_1^\mathrm{eff}/\calA$& $\alpha$& $\alpha/\calC$
\\\hline
circle & $\gamma_1^2/4\simeq 1.45$$^{a,b}$ & $4/\gamma_1^2\simeq 0.69$$^{a,b}$  &  $4\pi$& 2$^c$\\
quarter-circle  & 1.27$^d$ & 0.65$^d$  & $29.97$$^d$ & 1.85$^d$ \\
half-circle     & 1.38$^d$ & 0.64$^d$ & $33.17$$^d$& 1.97$^d$ \\
ellipse(1:2)    & 1.50$^d$ & 0.67$^d$ & $10\pi$$^c$& 2.10$^d$\\
ellipse(1:3)    & 1.54$^d$ & 0.62$^d$  & $40\pi/3$$^c$& 2.21$^d$\\
ellipse(1:4)    & 1.57$^d$ & 0.58$^d$  & $17\pi$$^c$&
2.28$^d$\\\hline
triangle(1:1:1) & $\pi^2/9\simeq 1.10$$^e$& $6/\pi^2\simeq 0.61$$^e$  & $20\sqrt{3}\:$$^c$ & $5/3\simeq 1.67$$^c$\\
triangle(1:1:$\sqrt{2}$) & $\frac{5\pi^2}
  {4( 2 + \sqrt{2})^2}\simeq 1.06$$^a$ &$512/9\pi^4\simeq 0.58$$^a$
& $38.33$$^d$&1.64$^d$\\\hline
square(1:1) & $\pi^2/8\simeq 1.23$$^a$ & $64/\pi^4\simeq 0.66$$^a$  & $28.45$$^d$ &  1.78$^d$\\
rectangle(1:2) & $5\pi^2/36\simeq 1.37$$^a$& $64/\pi^4\simeq 0.66$$^a$  &  $34.98$$^d$& 1.94$^d$\\
rectangle(1:3) & $ 5\pi^2/32\simeq 1.54$$^a$ & $64/\pi^4\simeq 0.66$$^a$ &  $45.57$$^d$& 2.14$^d$\\
rectangle(1:4) & $17\pi^2/100\simeq 1.68$$^a$ &$64/\pi^4\simeq 0.66$$^a$ & $56.98$$^d$ &  2.28$^d$ \\
rectangle(1:$\infty$) & $ \sim\pi^2/4\simeq 2.47$$^a$
&$64/\pi^4\simeq 0.66$$^a$ & $\infty$ & $\sim 3$$^f$\\\hline
pentagon &  1.30$^d$ & 0.67$^d$& $26.77$$^d$ &1.84$^d$\\\hline
hexagon& 1.34$^d$ & 0.68$^d$ & $26.08$$^d$& 1.88$^d$\\\hline
\end{tabular}
\caption{Central dimensionless parameters for different
geometries.
\newline $^a$See e.g.~\cite{Morse:1953} for the eigenmodes and
eigenspectrum. \newline $^b$Here, $\gamma_{1}\simeq 2.405$ is the
first root of the zeroth Bessel function of the first kind.
\newline $^c$See~\cite{Mortensen:05b}.\newline $^d$Data obtained by
finite-element simulations~\cite{comsol}.\newline $^e$See
e.g.~\cite{Brack:1997} for the eigenmodes and
eigenspectrum.\newline $^f$See e.g.~\cite{Batchelor:67} for a
solution of the Poisson equation.} \label{tab:1}
\end{center}
\end{table}

\subsubsection{Non-overlapping, thin Debye layers.}

For the off-diagonal elements of $G$ we use that
$[1+(\kappa_n\lambda_D)^2]^{-1}= 1 +{\cal O}[k_n^2(\lambda_D
\calP/\calA)^2]$. In Section~\ref{sec:numerics} we numerically
justify that the smallest dimensionless eigenvalue $k_1^2$ is of
the order unity, so we may approximate the sum by a factor of
unity, see Table~\ref{tab:1}, whereby we arrive at
Eq.~(\ref{eq:G12thin}) for $\lambda_D\ll \calA/\calP$. These
results for the off-diagonal elements are fully equivalent to the
Helmholtz--Smoluchowski result~\cite{Probstein:94a}. For $G_{22}$
we use that $(\kappa_n\lambda_D)^2[1+(\kappa_n\lambda_D)^2]^{-2}=
{\cal O}[k_n^2(\lambda_D \calP/\calA)^2]$, thus we may neglect the
second term, whereby we arrive at Eq.~(\ref{eq:G22thin}).

\subsubsection{Strongly overlapping Debye layers.}
\label{sec:DH_strongoverlap}

In the case of $\kappa_1\lambda_D\gg 1$ we may use the result
$[1+(\kappa_n\lambda_D)^2]^{-1}=(\kappa_n\lambda_D)^{-2}+{\cal
O}[k_n^{-4}(\lambda_D \calP/\calA)^{-4}]$ which gives
\begin{equation}\label{eq:G12_strong_DH}
G_{12}=G_{21}\simeq
-\frac{\epsilon\zeta}{\lambda_D^2}\:G_{11},\quad \lambda_D\gg
\frac{\calA}{\calP}.
\end{equation}
This is the Debye--H\"uckel limit of Eq.~(\ref{eq:G12strong}) for
strongly overlapping Debye layers. For $G_{22}$ we use
Eq.~(\ref{eq:G22_G12}) and arrive at the result in
Eq.~(\ref{eq:G22strong}) for $Ze\zeta\ll k_BT$ by using
$\sigma_\mathrm{\!o}=\epsilon D/\lambda_D^2$.

\subsubsection{The circular case.}

For a circular cross-section, where $\alpha=8\pi$, it can be shown
that~\cite{Probstein:94a}
 \begin{equation}\label{eq:G12circle}
 G_{12}^\mathrm{circ}
 =G_{21}^\mathrm{circ}=-\frac{\calA}{L}\frac{\epsilon\zeta}{\eta}
 \frac{I_2\!\big(\sqrt{8\calA/\alpha\lambda_D^2}\:\big)}{I_0\!\big(\sqrt{8\calA/\alpha\lambda_D^2}\:\big)},
 \end{equation}
where $I_n$ is the $n$th modified Bessel function of the first
kind, and were we have explicitly introduced the variable
$\calA/\alpha\lambda_D^2$ to emphasize the asymptotic dependence
in Eq.~(\ref{eq:G12strong}) for strongly overlapping Debye layers.
We note that we recover the limits in Eqs.~(\ref{eq:G12thin}) and
(\ref{eq:G12strong}) for $\lambda_D\ll \calA/\calP$ and
$\lambda_D\gg\calA/\calP$, respectively.

\section{Beyond the Debye--H{\"u}ckel approximation}\label{sec:BeyondDebyeHuckel}

In order to go beyond the Debye--H\"uckel approximation we
consider, for simplicity, a symmetric binary $(Z$:$Z)$
electrolyte. Next, we neglect strong correlations between the ions
so that the equilibrium properties of the electrolyte are governed
by Boltzmann statistics~\cite{Squires:05a}, i.e., the
concentrations of the two type of ions are given by
 \begin{equation}\label{eq:c:boltzmann}
 c_\mathrm{eq}^\pm(\rrrp) = c_\mathrm{o}\exp\!\left[\mp\frac{Ze}{k_BT}\:\phi_\mathrm{eq}(\rrrp)\right].
 \end{equation}
This is equivalent to assuming equilibrium with bulk reservoirs
at the ends of the channel in which the potential
$\phi_\mathrm{eq}$ tends to zero and both concentrations
$c_\mathrm{eq}^\pm$ to $c_\mathrm{o}$.

Substituting the charge density
$\rho_\mathrm{eq}^e=Ze(c_\mathrm{eq}^+ - c_\mathrm{eq}^-)$ into
the Poisson equation (\ref{eq:P}) we arrive at the nonlinear
Poisson--Boltzmann equation~\cite{Squires:05a,Probstein:94a}
 \begin{equation}
 \Laplace \phi_\mathrm{eq}(\rrrp)= \frac{k_BT}{Ze\,\lambda_D^2}\sinh\!\left[\frac{Ze}{k_BT}\:\phi_\mathrm{eq}(\rrrp)\right],
 \label{eq:poissonboltzmann}
 \end{equation}
where the Debye screening length is given by
\begin{equation}
\lambda_D= \sqrt{\frac{\epsilon k_B T}{2 (Ze)^2 c_\mathrm{o}}}.
\end{equation}
The conductivity $\sigma$ of the electrolyte depends on the local
ionic concentrations
 \begin{equation}\label{eq:sigma:nonlin}
 \sigma(\rrrp) = \frac{(Ze)^2 D}{k_BT}
 \big[c_\mathrm{eq}^+(\rrrp) + c_\mathrm{eq}^-(\rrrp)\big]
 = \sigma_\mathrm{\!o}\cosh\!\left[\frac{Ze}{k_BT}\:\phi_\mathrm{eq}(\rrrp)\right],
 \end{equation}
assuming equal diffusivities $D$ for the two ionic species. In the
Debye--H\"uckel limit, $Ze \zeta \ll k_BT$, where thermal energy
dominates over the electrostatic energy we may linearize the
right-hand side of Eq.~(\ref{eq:poissonboltzmann}) so that we
arrive at the Debye--H\"uckel approximation in
Eq.~(\ref{eq:PB_DH}). Similarly the conductivity in
Eq.~(\ref{eq:sigma:nonlin}) reduces to the bulk conductivity
$\sigma_\mathrm{\!o}$. Compared to Eq.~(\ref{eq:I}) the electrical
current obtained from the nonlinear Nernst--Planck equation
becomes
 \begin{equation}\label{eq:current:nonlin}
 I = \big<\rho_\mathrm{eq}^e\big|v\big> + E\big<\sigma\big|1\big>.
 \end{equation}
We calculate the off-diagonal elements from
$G_{12}=G_{21}=\big<\rho_\mathrm{eq}^e\big|v_p\big>/\Delta p$ and find
 \begin{equation}\label{eq:G21:nonlin}
 G_{12}=G_{21} = -\frac{1}{\Delta p}\frac{\epsilon k_BT}{Ze\lambda_D^2}
 \Big<\sinh\!\Big(\frac{Ze}{k_BT}\:\phi_\mathrm{eq}\Big)\Big|v_p\Big>.
 \end{equation}
Similarly, Eq.~(\ref{eq:G22_general}) for the two components in
the electrical conduction
$G_{22}=G_{22}^\mathrm{mig}+G_{22}^\mathrm{conv}$ we get
\begin{eqnarray}
 G_{22}^\mathrm{mig}
&=&
\frac{\sigma_\mathrm{\!o}}{L}\Big<\cosh\!\Big(\frac{Ze}{k_BT}\:\phi_\mathrm{eq}\Big)\Big|
 1\Big>,\label{eq:G22:nonlin:mig}\\
 G_{22}^\mathrm{conv}
&=&  \frac{\sigma_\mathrm{\!o} m }{2L}
 \frac{Ze}{k_BT}\Big<\sinh\!\Big(\frac{Ze}{k_BT}\:\phi_\mathrm{eq}\Big)\Big|
 \zeta-\phi_\mathrm{eq}\Big>, \label{eq:G22:nonlin:conv}
 \end{eqnarray}
where we have used that $\sigma_\mathrm{\!o}=\epsilon
D/\lambda_D^2$ and introduced the dimensionless quantity $m$,
\begin{equation}
m\equiv\left(\frac{k_BT}{Ze}\right)^2\frac{2\epsilon}{\eta
 D},\label{eq:m}
\end{equation}
which indicates the importance of electro-osmosis relative to
electro-migration.

\subsection{Non-overlapping, thin Debye layers}
In the limit of thin Debye layers we have already discussed how
Eq.~(\ref{eq:G12_general}) in general leads to
Eq.~(\ref{eq:G12thin}) because the screening is good and
$\phi_\mathrm{eq}$ is nonzero only on a negligible part of
$\Omega$. This property is more implicit when $G_{12}$ or $G_{21}$
is written in the form of Eq.~(\ref{eq:G21:nonlin}), which is a
more appropriate starting point for analyzing the limit of
strongly overlapping Debye layers. For $G_{22}$ the calculations
are more involved; we assume that the channel wall is sufficiently
smooth on the Debye-length scale so that we can everywhere use the
Gouy--Chapman (GC) solution for a semi-infinite planar
geometry~\cite{Squires:05a,Probstein:94a},
 \begin{equation}\label{eq:phieq:gouychapman}
 \phi_\mathrm{eq}^\mathrm{GC} =
 \frac{k_BT}{Ze}\,4\tanh^{-1}\!\Big[\tanh\!\bigg(\frac{Ze\,\zeta}{4k_BT}\bigg)
 e^{-r_n/\lambda_D}\Big],
 \end{equation}
where $r_n$ denotes the normal distance to the channel wall.
Substituting this into Eq.~(\ref{eq:G22:nonlin:mig}) and
(\ref{eq:G22:nonlin:conv}) the integrals can be carried out
analytically resulting in
 \begin{equation}\label{eq:G22thin:nonlin}
 G_{22} = \sigma_\mathrm{\!o}\frac{\calA}{L}\big(1 + Du\big), \quad \lambda_D\ll \frac{\calA}{\calP},
 \end{equation}
where $Du$ is the Dukhin number
 \begin{equation}\label{eq:dukhin}
 Du = \frac{\lambda_D\calP}{\calA}
 (1+m)
 \:4\sinh^2\!\bigg(\frac{Ze\zeta}{4k_BT}\bigg),
 \end{equation}
defined as the ratio of the surface conductivity in the charged
Debye layers to the bulk conductivity $\sigma_\mathrm{\!o}$ times
the geometrical length scale $\calA/\calP$~(see
Ref.~\cite{Bazant:04a} and references therein). Clearly, when the
Debye layer becomes very thin, surface conduction is negligible
and we recover the simple result in Eq.~(\ref{eq:G22thin}).

\subsection{Strongly overlapping Debye layers}
\label{sec:nonlin_strongoverlap}
When the Debye layers are strongly overlapping the screening is
weak and $\phi_\mathrm{eq}\approx\zeta$ throughout the cross section.
Hence we can pull the integrand $\sinh(Ze\phi_\mathrm{eq}/k_BT)$
outside the \emph{bra-ket} in Eq.~(\ref{eq:G21:nonlin}) and we
arrive at Eq.~(\ref{eq:G12strong}).
Here, we have used that $\langle1|v_p\rangle/\Delta p = G_{11}$
and introduced the parameterization in Eq.~(\ref{eq:G11_alpha}).
Similarly, from Eqs.~(\ref{eq:G22:nonlin:mig}) and
(\ref{eq:G22:nonlin:conv}) we obtain Eq.~(\ref{eq:G22strong})
where we have used Eqs.~(\ref{eq:G12_general}) and
(\ref{eq:G12strong}) to eliminate
$\langle1|\zeta-\phi_\mathrm{eq}\rangle$.

We note that due to shifts in free energies, the zeta potential
inside a narrow channel with significant Debye-layer overlap is
generally not the same as in a macroscopic channel with no
overlap, see e.g.~\cite{Vanderheyden:2005,Behrens:01} for a
discussion.

\begin{figure}[b!]
\begin{center}
\epsfig{file=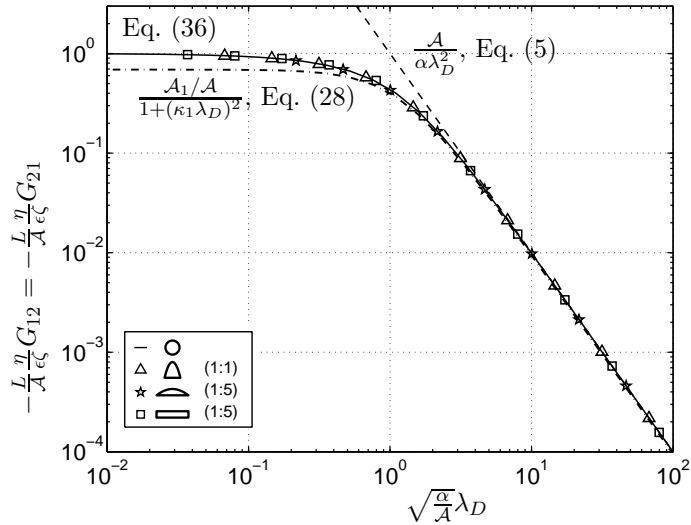, width=0.7\columnwidth, clip}
\end{center}
\caption{Rescaled off-diagonal transport coefficients versus
rescaled Debye-layer thickness in the Debye--H\"uckel limit. The
solid line is the exact result for a circle,
Eq.~(\ref{eq:G12circle}), and the dashed line shows
Eq.~(\ref{eq:G12strong}) for $Ze\zeta\ll k_BT$. The data points
are finite-element simulations in the linearized regime for
different cross sections, see inset. Finally, the dash-dotted line
shows the first term from the summation in
Eq.~(\ref{eq:G12general}) only. \label{fig2}}
\end{figure}

\begin{figure}[t!]
\begin{center}
\epsfig{file=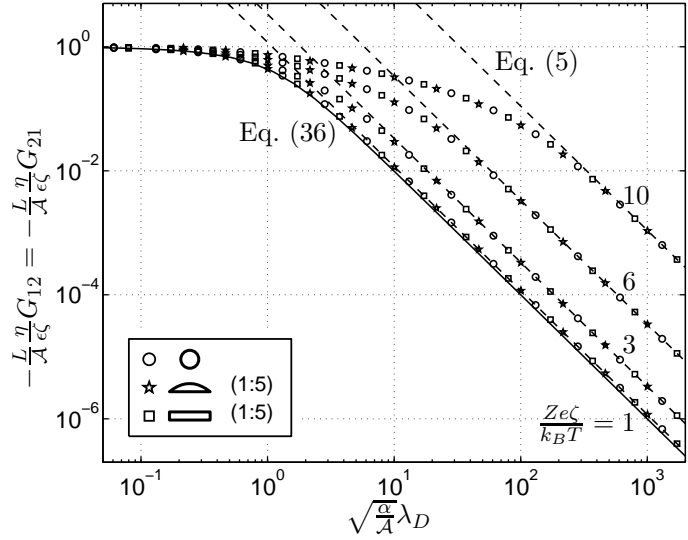, width=0.7\columnwidth, clip}
\end{center}
\caption{Rescaled off-diagonal transport coefficients versus
rescaled Debye-layer thickness beyond the Debye--H\"u{}ckel
approximation. The solid line is the exact result for a circle
within the Debye--H\"u{}ckel approximation,
Eq.~(\ref{eq:G12circle}). The data points are finite-element
simulations for different cross sections (see inset) with
$Ze\zeta/k_BT=1$, $3$, $6$, and $10$ from below. The dashed lines
indicate the corresponding asymptotic expression for strong
Debye-layer overlap, Eq.~(\ref{eq:G12strong}).\label{fig3}}
\end{figure}

\begin{figure}[t!]
\begin{center}
\epsfig{file=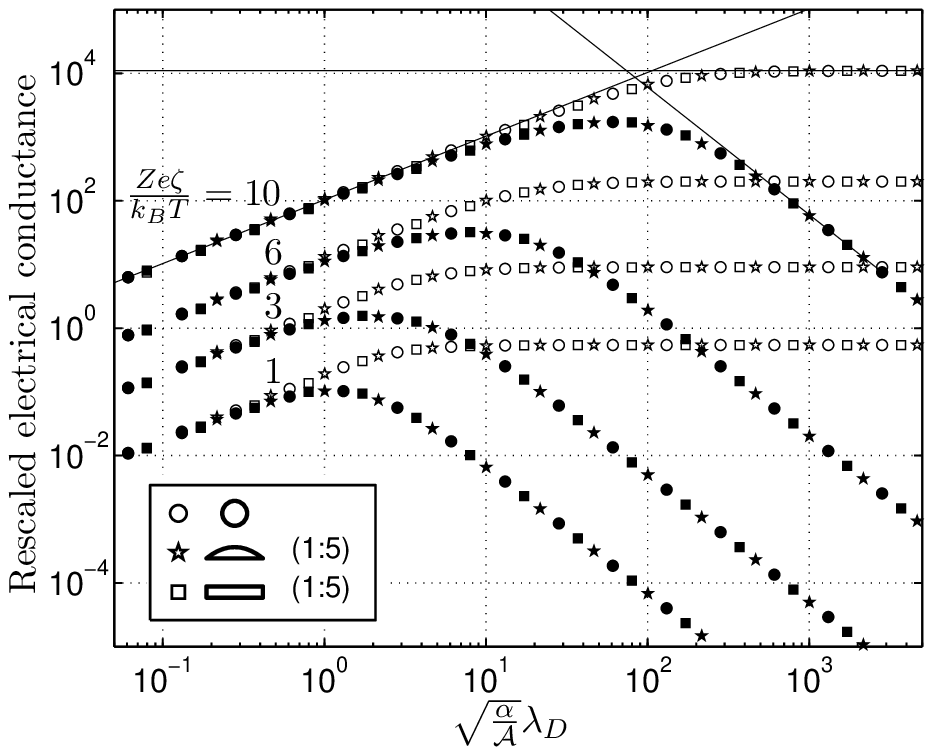, width=0.7\columnwidth, clip}
\end{center}
\caption{Comparison of the two components in the $G_{22}$
transport coefficient for three different geometries, see inset.
Open symbols show $(G_{22}^\mathrm{mig}L/
\sigma_\mathrm{\!o}\calA) - 1$, i.e., the surface specific
contribution to the electrical conductance from electro-migration,
see Eq.~(\ref{eq:G22:nonlin:mig}). Solid symbols show the
contribution from electro-osmotic convection
$G_{22}^\mathrm{conv}L/m\sigma_\mathrm{\!o}\calA$, see
Eq.~(\ref{eq:G22:nonlin:conv}); for ease of comparison we have
included the dimensionless number $m$ in the rescaling. The data
are obtained from finite-element simulations for different cross
sections (see inset) with $Ze\zeta/k_BT=1$, $3$, $6$, and $10$.
The solid lines indicate the corresponding asymptotic expressions
for strong and weak Debye-layer overlap, Eqs.~(\ref{eq:G22strong})
and (\ref{eq:G22thin:nonlin}), respectively, for $Ze\zeta/k_BT=10$
in a circular channel. Asymptotes for the other values of
$Ze\zeta/k_BT$ have been left out for clarify, but equally good
agreement is found also in these cases.\label{fig4}}
\end{figure}

\section{Numerical results}
\label{sec:numerics}

\subsection{The Helmholtz basis}

Only few geometries allow analytical solutions of both the
Helmholtz equation and the Poisson equation. The circle is of
course among the most well-known solutions and the equilateral
triangle is another example. However, in general the equations
have to be solved numerically, and for this purpose we have used
the commercially available finite-element software
Femlab~\cite{comsol}. The first eigenstate of the Helmholtz
equation is in general non-degenerate and numbers for a selection
of geometries are tabulated in Table 1. Note how the different
numbers converge when going through the regular polygons starting
from the equilateral triangle through the square, the regular
pentagon, and the regular hexagon to the circle. In general,
$k_1^2$ is of the order unity, and for relevant high-order modes
(those with a nonzero $\calA_n$) the eigenvalue is typically much
larger. Similarly, for the effective area we find that
$\calA_1/\calA\leq 4/\gamma_1^2\simeq 0.69$ and consequently we
have $\calA_n/\calA< 1- 4/\gamma_1^2\simeq 0.31$ for $n\geq 2$.
The transport coefficients in Eqs.~(\ref{eq:Ggeneral})
to~(\ref{eq:Ggeneral:4}) are thus strongly influenced by the first
eigenmode which may be used for approximations and estimates of
the transport coefficients. As an example the column for
$\alpha/\calC$ is well approximated by only including the first
eigenvalue in the summation in Eq.~(\ref{eq:alpha}).

\subsection{Transport coefficients}

Our analytical results predict that when going to either of the
limits of thin non-overlapping or strongly overlapping Debye
layers, the transport coefficients only depend on the channel
geometry through the cross sectional area $\calA$ and the
correction factor $\alpha$. Therefore, when plotted against the
rescaled Debye length $\sqrt{\alpha/\calA}\:\lambda_D$, all our
results should collapse on the same asymptotes in the two limits.

In Fig.~\ref{fig2} we show the results for the off-diagonal
coefficients obtained from finite-element simulations in the
Debye--H\"uckel limit $Ze\zeta\ll k_BT$ for three different
channel cross sections, namely two parabola shaped channels of
aspect ratio 1:1 and 1:5, respectively, and a rectangular channel
of aspect ratio 1:5. In all cases we find excellent agreement
between the numerics and the asymptotic expressions. For the
comparison we have also included exact results,
Eq.~(\ref{eq:G12circle}), for the circular cross section as well
as results based on only the first eigenvalue in
Eq.~(\ref{eq:G12general}). Even though Eq.~(\ref{eq:G12circle}) is
derived for a circular geometry we find that it also accounts
remarkably well for even highly non-circular geometries in the
intermediate regime of weakly overlapping Debye layers.

In Fig.~\ref{fig3} we show numerical results for the off-diagonal
transport coefficients beyond the Debye--H\"uckel approximation.
At large zeta potentials the Debye layer is strongly compressed
and the effective screening length reduced. Therefore the
suppression of the electro-osmotic flow/streaming current at
strong Debye-layer overlap is shifted to larger values of
$\lambda_D$ as compared to the Debye--H\"uckel limit. For the
comparison we have also included the exact result for a circular
cross section in the Debye--H\"uckel approximation as well as the
asymptotic expression for non-overlapping and strongly overlapping
Debye layers, Eq.~(\ref{eq:G12strong}). As seen the asymptotic
expressions account well for the full numerical solutions
independently of the geometry.

In Fig.~\ref{fig4} we show numerical results for the electrical
conductance beyond the Debye--H\"uckel approximation. Open symbols
show the electro-migration part $G_{22}^\mathrm{mig}$, subtracted
the trivial bulk contribution $\sigma_\mathrm{\!o}\calA/L$,
whereas solid symbols show the electro-osmotic convection part
$G_{22}^\mathrm{conv}$, see Eqs.~(\ref{eq:G22:nonlin:mig})
and~(\ref{eq:G22:nonlin:conv}), respectively. Again, we find that
the numerics are in excellent agreement with our asymptotic
results. For the $\lambda_D\ll\calP/\calA$ regime we note that the
Dukhin number, Eq.~(\ref{eq:dukhin}), is proportional to
$\lambda_D\calP/\calA=\sqrt{\calC/\calA}\:\lambda_D$ and not
$\sqrt{\alpha/\calA}\:\lambda_D$. Therefore, strictly one would
not in general expect data to collapse on the same asymptote for
$\lambda_D\ll\calP/\calA$. Looking carefully at this part of the
figure, small variations can be seen from geometry to geometry --
the reason why the variations are still so small is that
$\alpha/\calC\sim 2$ independently of geometry, see Table~1.

\section{Conclusion}

We have analyzed the flow of incompressible electrolytes in long,
straight channels driven by pressure and electro-osmosis. By using
a powerful Hilbert space eigenfunction expansion we have been able
to address the general problem of an arbitrary cross section and
obtained general results for the hydraulic and electrical
transport coefficients. Results for strongly overlapping and thin,
non-overlapping Debye layers are particular simple, and from these
analytical results we have calculated the corrections to the
hydraulic resistance due to electro-hydrodynamic interactions.
These analytical results reveal that the geometry dependence only
appears through the area $\calA$ and the correction factor
$\alpha$, as the expressions only depend on the rescaled Debye
length $\sqrt{\alpha/\calA}\:\lambda_D$. Our numerical analysis
based on finite-element simulations indicates that these
conclusions are generally valid also for intermediate values of
$\lambda_D$. Combined with recent detailed work on the geometrical
correction factor~\cite{Mortensen:05b} the present results
constitute an important step toward circuit
analysis~\cite{ajdari:04a} of complicated micro and nanofluidic
networks incorporating complicated cross-sectional channel
geometries.

\vspace{10mm}


\end{document}